\documentclass[]{spie}  

\usepackage{amsmath,amsfonts,amssymb}
\usepackage{graphicx}
\usepackage[colorlinks=true, allcolors=blue]{hyperref}
\usepackage{color}

\title{The Keck Planet Imager and Characterizer:
Demonstrating advanced exoplanet characterization techniques for future extremely large telescopes}

\author{N.~Jovanovic$^{a}$, J. R.~Delorme$^{a}$, C. Z. Bond$^{c,d}$, S. Cetre$^{c}$, D.~Mawet$^{a,b}$, D.~Echeverri$^{a}$, J. K.~Wallace$^{b}$, R. Bartos$^{b}$, S. Lilley$^{c}$, S. Ragland$^{c}$, G.~Ruane$^{b}$, P. Wizinowich$^{c}$, M. Chun$^{d}$, J. Wang$^{a}$, J. Wang$^{f}$, M. Fitzgerald$^{e}$, K. Matthews$^{a}$, J. Pezzato$^{a}$, B. Calvin$^{a}$, M. Millar-Blanchaer$^{a,b}$, E. C. Martin$^{g}$, E. Wetherell$^{c}$, E. Wang$^{e}$, S. Jacobson$^{d}$, E. Warmbier$^{d}$, C. Lockhart$^{d}$, D. Hall$^{d}$, R. Jensen-Clem$^{h}$ and E. McEwen$^{h}$

$^{a}$ Department of Astronomy, California Institute of Technology, 1200 E. California Blvd., Pasadena, CA, 91125, USA; \\
$^{b}$ Jet Propulsion Laboratory, California Institute of Technology, 4800 Oak Grove Drive, Pasadena, CA, 91109, USA; \\
$^{c}$ W. M. Keck Observatory, 65-1120 Mamalahoa Hwy., Kamuela, HI, 96743, USA; \\
$^{d}$University of Hawaii, 640 N. Aohoku Place, Hilo, HI, USA 96720; \\
$^{e}$Department of Physics and Astronomy, University of California-Los Angeles, 430 Portola Plaza, Los Angeles, CA, USA 90095;\\
$^{f}$Department of Astronomy, The Ohio State University, 100 W 18th Ave, Columbus, OH, USA 43210, USA;\\
$^{g}$Department of Astronomy \& Astrophysics, University of California, Santa Cruz, CA 95064, USA;\\
$^{h}$University of California, Berkeley, 510 Campbell Hall, Astronomy Department, Berkeley, CA 94720, USA;\\
}

\authorinfo{Further author information: (Send correspondence to Nemanja Jovanovic)\\Nemanja Jovanovic: E-mail:  jovanovic.nem@gmail.com, Telephone: +1 626 395 1214}

\begin{document} 
\maketitle

\begin{abstract}
The Keck Planet Imager and Characterizer (KPIC) is an upgrade to the Keck II adaptive optics system enabling high contrast imaging and high-resolution spectroscopic characterization of giant exoplanets in the mid-infrared (2-5 microns). The KPIC instrument will be developed in phases. Phase I entails the installation of an infrared pyramid wavefront sensor (PyWFS) based on a fast, low-noise SAPHIRA IR-APD array. The ultra-sensitive infrared PyWFS will enable high contrast studies of infant exoplanets around cool, red, and/or obscured targets in star forming regions. In addition, the light downstream of the PyWFS will be coupled into an array of single-mode fibers with the aid of an active fiber injection unit (FIU). In turn, these fibers route light to Keck's high-resolution infrared spectrograph NIRSPEC, so that high dispersion coronagraphy (HDC) can be implemented for the first time. HDC optimally pairs high contrast imaging and high-resolution spectroscopy allowing detailed characterization of exoplanet atmospheres, including molecular composition, spin measurements, and Doppler imaging.

Here we provide an overview of the instrument, its science scope, and report on recent results from on-sky commissioning of Phase I. The instrument design and techniques developed will be key for more advanced instrument concepts needed for the extremely large telescopes of the future. 
\end{abstract}


\section{INTRODUCTION}
\label{sec:intro}  
High contrast imaging is a crucial tool for the characterization of exoplanets that can be used to constrain the luminosity, orbit, chemical composition and abundances of exoplanets. The Keck Planet Imager and Characterizer (KPIC) is designed to enhance the high contrast imaging capabilities of the Keck II telescope~\cite{Mawet2016_KPIC}. KPIC is specifically engineered to promote two primary science cases: 1) the study of infant exoplanets around cool, red, and/or obscured targets in star forming regions and 2) the detailed spectroscopic characterization of such targets. 

\begin{figure} [b!]
   \begin{center}
   \begin{tabular}{c} 
   \includegraphics[width=0.98\textwidth]{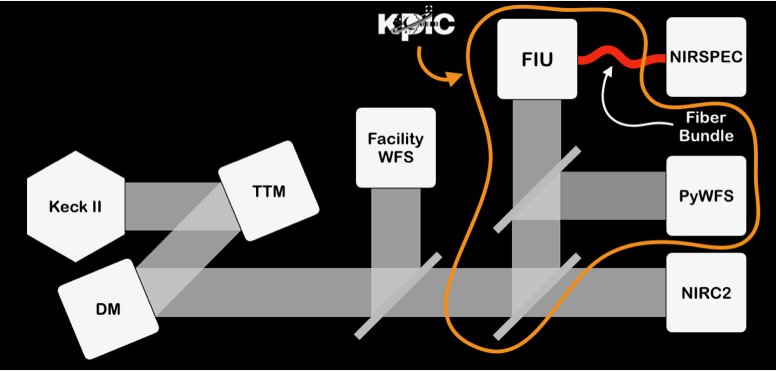}
   \end{tabular}
   \end{center}
   \caption[example] 
   { \label{fig:KPIC} 
A block diagram of Keck AO system including the KPIC upgrades (circled in orange) as well as NIRC2 and NIRSPEC.}
\end{figure}

KPIC enables these science cases through a series of strategic upgrades to the Keck II adaptive optics (AO) system, NIRC2 and NIRSPEC instruments. Figure~\ref{fig:KPIC} shows the location of the upgrades in Keck AO. The upgrades can be broken into several themes, which we outline here. 

\subsection{Advanced wavefront control}
Key to obtaining high contrast images on-sky, is the ability to restore the PSF to high Strehl and reduce residual star light contamination across the focal plane. This can be achieved through wavefront control in a number of ways that KPIC aims to implement.  

KPIC improves on the wavefront control provided by the Keck AO system by implementing a pyramid wavefront sensor (PyWFS)~\cite{bond2018,cetre2018}. Besides the fact that a PyWFS has a greater sensitivity than a Shack-Hartmann, there are several other features of the KPIC implementation worth mentioning. Firstly, the KPIC PyWFS operates in the H-band. By operating at NIR wavelengths, the fractional wavefront error is lower compared to a visible wavefront sensor (WFS) and so the PyWFS operates in the linear regime even with poorer conditions. Secondly, the sensor utilizes a SAPHIRA avalanche photodiode array, which enables the possibility for sub-electron read noise at kHz frame rates. Thridly, the PyWFS utilizes two roof prisms to mimic the effect of a single pramidal prism. This was first employed on SCExAO and is a simple and cost effective solution for realizing a PyWFS~\cite{lozi2019}. All of these features combine to enable a highly sensitive WFS that will be used to improve the Strehl ratio compared to the facility Shack-Hartmann, ideal for direct imaging of exoplanets. The PyWFS will initially drive the 349 actuator Xinetics DM in Keck AO, which is common path to the NIRC2 instrument as well as the new fiber injection unit (FIU) that will be discussed below. 

In addition to a new WFS, a 1000 element deformable mirror (DM) will also be deployed. This DM will be located after the KPIC pickoff and hence only common to the PyWFS and the FIU (not NIRC2). The Boston MicroMachines (BMC) DM will allow for correction of higher spatial frequency modes than the Xinetics, increasing the control radius in the focal plane and ultimately improving both the Strehl and the contrast.   

Besides the addition of a new WFS and a DM, new control algorithms will also be implemented. Predictive control techniques will be applied to overcome the latency in the control loop~\cite{eden2019}. The approach used follows that outlined by Guyon and Males (2017)~\cite{guyon2017,males2018}, which could offer significant improvements in performance. Focal plane wavefront control is another avenue that KPIC will explore. This will be done in the form of enhancing the existing speckle nulling routine that operates in conjunction with NIRC2 as well as enabling speckle suppression through the single mode fiber used by NIRSPEC, discussed in the next section.  

\subsection{Coronagraph improvements}
\begin{figure} [b!]
   \begin{center}
   \begin{tabular}{c} 
   \includegraphics[width=0.98\textwidth]{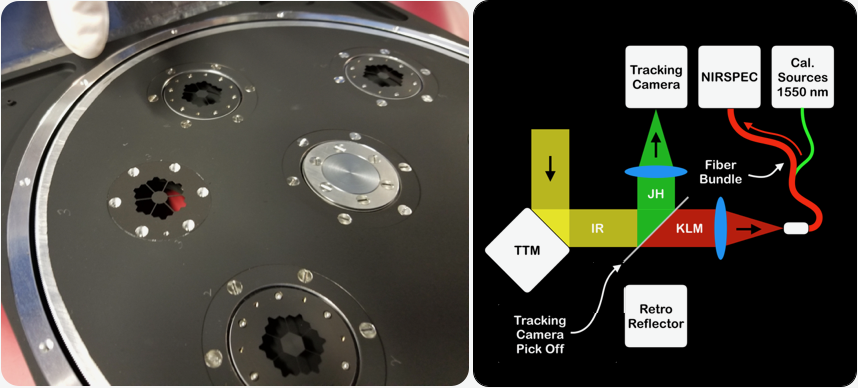}
   \end{tabular}
   \end{center}
   \caption[example] 
   { \label{fig:lyot} 
(Left) Image of Lyot stops installed in the filter wheel inside NIRC2. (Right) A schematic of the key fiber injection unit components.}
\end{figure}

The NIRC2 imager is unique amongst high contrast imaging platforms as it offers imaging out to M band. At L and M band, the Keck AO system (Shack-Hartmann+Xinetics DM) achieve Strehl ratios $>80\%$ and fall into the so called ``extreme AO" regime. This has been extensively exploited for the direct imaging of young giant planets and disks~\cite{Mawet2017b,mawet2019,ruane2019,currie2019}. The KPIC project enhances this capability by virtue of introducing vortex phase masks optimized specifically for K, L and M bands individually to replace the single L band mask used in the past. In addition, several new Lyot stops (shown in the left panel of Fig.~\ref{fig:lyot}) have been introduced that are less conservative than previous masks increasing planet throughput while matching the Keck pupil more accurately, reducing stellar leakage. 

Finally, a polarimetric mode will be added to NIRC2. This will consist of a Wollaston prism installed inside the filter wheels in NIRC2 to create two orthogonally polarized beams on the detector. A half waveplate (HWP) will be placed at the input to Keck AO to modulate the polarization of the beams on the detector. 

\subsection{High resolution spectroscopy}
High dispersion coronagraphy (HDC) is a technique that allows for the spectral characterization of an exoplanet atmosphere at resolutions much higher than a integral field spectrograph can achieve, while offering superior stellar suppression~\cite{snellen2015,mawet2017}. The technique relies on using a single mode fiber in a focal plane downstream of a coronagraph to send the known planets light to a high resolution spectrograph to study its properties~\cite{wang2017}. The extracted spectrum is then cross-correlated with template spectra of the molecular species one might expect to see in the atmosphere of the planet, to infer the presence and possibly the abundance of that species.   

KPIC will enable HDC for the first time. Given the coronagraphs in NIRC2 are in a cryostat, there is currently no way to build a fiber injection post coronagraph in NIRC2 (without a serious overhaul). Therefore the FIU is being developed after the KPIC pickoff in the arm of the PyWFS. 

A schematic of the key components of the FIU is shown in the right panel of Fig.~\ref{fig:lyot}. The K and L band light is transmitted to a focal plane where a single mode fiber (SMF) bundle is located. The J and H band light remaining after the PyWFS pickoff is reflected off a dichroic to a tracking camera (FirstLight Imaging, C-red2). The tacking camera is used to monitor the position of the target and drive the tip/tilt mirror (TTM) to precisely position it on the SMF. The light coupled into the SMF is routed to NIRSPEC, a high resolution spectrograph that operates from y-M band with a resolving power as high as R$\sim37000$. NIRSPEC was originally designed to be a seeing-limited spectrograph so fore-optics known as the fiber extraction unit (FIU), reimage and magnify the output of the SMFs onto the slit of NIRSPEC in order to make it work with a diffraction limited feed. 

To enhance the starlight suppression for the FIU arm and hence really demonstrate HDC, the 1000 element DM, as well as several coronagraphs~\cite{echeverri2019}, an atmospheric dispersion compensator and pupil shaping optics will be installed upstream of the fiber feed during a later phase of the project~\cite{pezzato2019}.

\section{Development strategy and status}~\label{sec:development}
The KPIC instrument is being deployed as a PI instrument at first to Keck and will be facilitized at a later date. To expedite deployment, the project is broken into phases. The phases are summarized below:

\begin{itemize}
    \item {Phase I: The NIR PyWFS and the base FIU optics. The base FIU optics simply include the TTM, the tracking camera pickoff, the tracking camera, and the fiber coupling optics. In addition, the FEU was also installed into NIRSPEC. In phase I the K, L and M band coronagraphs were installed in NIRC2 along with the new Lyot stops and the Wollaston prism.}
    \item {Phase II: This involves replacing the base FIU optics with the same optical layout with other modules that improve planet throughput and suppress starlight to reduce the required integration time~\cite{pezzato2019,echeverri2019}. In addition, the PyWFS will be facilitized. This not only includes preparing the PyWFS for routine operation, but a replacement of the roof prisms for ones optimized for the NIR as well as steering capabilities on the input field steering mirrors to allow for offsetting for calibration. In parallel to the FIU and PyWFS upgrades, a HWP mechanism will be installed at the input of Keck AO to calibrate the new polarimetric mode of NIRC2.}
    \item{Phase III: Advanced wavefront control algorithms such as various forms of focal plane wavefront control, both in the focal plane of NIRC2 and through the fiber will be developed and tested to enhance contrast. In addition, the FIU will be facilitized.}
\end{itemize}

The phased deployment allows for the optics, mechanics and electronics to be sent to Keck early and fit tested and installed into Keck II AO. In addition, control software can begin integration with the Keck infrastructure. This helps arrest risk and gets the modules on-sky early on, enabling preliminary science, albeit with reduced capabilities. Later phases can therefore focus on optimizing performance without the concern for compliance with Keck interfaces. 

\begin{figure} [t!]
   \begin{center}
   \begin{tabular}{c} 
   \includegraphics[width=0.98\textwidth]{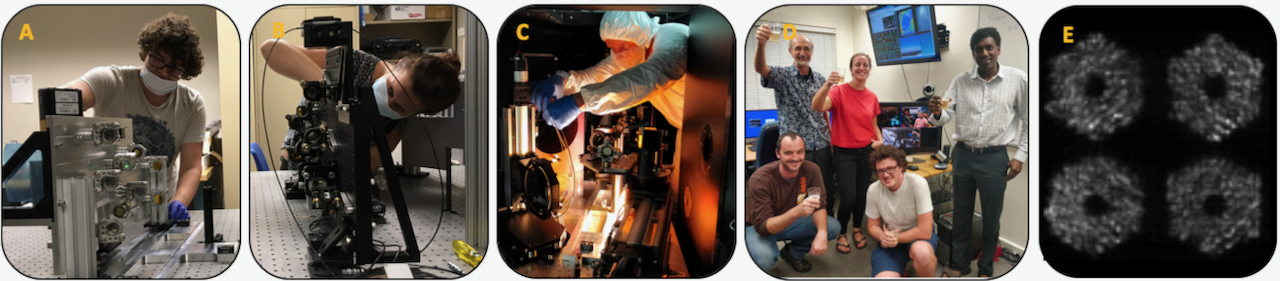}
   \end{tabular}
   \end{center}
   \caption[example] 
   { \label{fig:deploy} 
Images showing the development of KPIC Phase I at different stages. (a) Laboratory assembly and testing (Jacques-Robert Delorme shown aligning the FIU plate). (b) Full system integration in the laboratory (Charlotte Bond shown aligning the PyWFS plate) (c) Install in Keck AO (Nemanja Jovanovic shown installing cables in Keck AO). (d) The team celebrating the success of first light with the PyWFS. (e) An example PyWFS image of the 4 pupils with the loop closed on-sky.}
\end{figure}

The phase I hardware for the PyWFS, FIU and FEU was deployed to the Keck II telescope in September 2018 (see Fig.~\ref{fig:deploy} for a visual representation of the deployment process for phase I). The PyWFS started commissioning in late 2018 and is now fully operational. The FIU is still undergoing commissioning but is mostly operational and about to have its first multi-night science run in October 2019. The coronagraphic masks were deployed in March 2019 and after some optimization, are fully operational with the QACITS vortex alignment software~\cite{huby2017}. The Wollaston prism was deployed at the same time. 

Phase II is in the final stages of the design and planning. The FIU upgrades passed a detailed design review back in May 2019 and are now in the final stages of optimization before major procurement. We have initiated some orders on optics and mechanics to get them into the lab for early assembly and characterization to mitigate risk. All modules will be tested in a dedicated FIU testbed being replicated in the Exoplanet Technology Laboratory (ETL) at Caltech. Once verification and validation is complete, they will be shipped to be installed at Keck. Similarly, planning for the PyWFS facilitization is well underway and some parts have been ordered. The FIU will be shipped from Caltech to Keck in March 2020, for a summit install along with the PyWFS facilitization work in April 2020. The HWP for the polarimetric mode will likewise be deployed during this period. 

Phase III is currently in its infancy. There are several laboratory experiments underway in the ETL to test various focal plane wavefront sensing techniques for application in phase III. In addition, the control software for KPIC is being designed in phase I and II to be compliant with the more advanced needs of phase III.

\section{Preliminary on-sky results}~\label{sec:results}
\subsection{The NIR pyramid wavefront sensor}
\begin{figure} [t!]
   \begin{center}
   \begin{tabular}{c} 
   \includegraphics[width=0.98\textwidth]{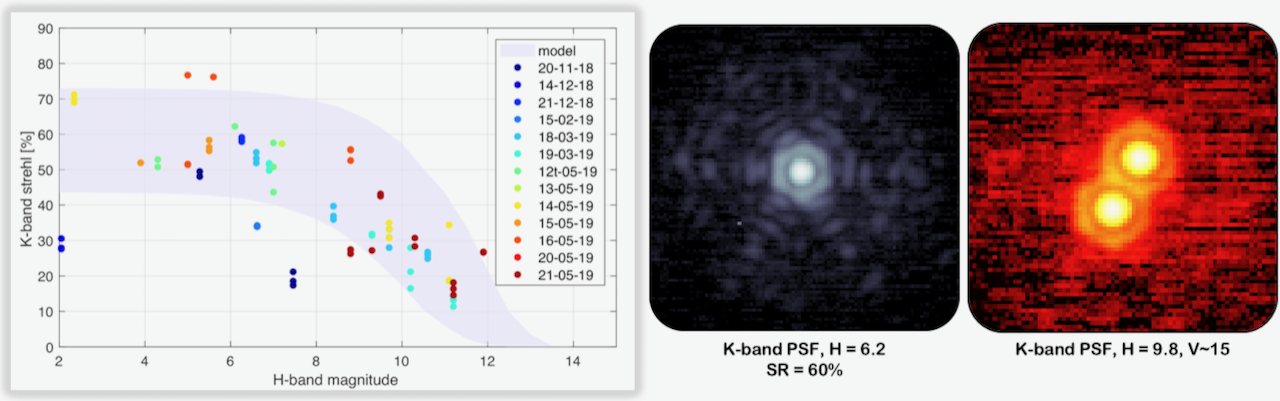}
   \end{tabular}
   \end{center}
   \caption[example] 
   { \label{fig:pywfsresults} 
(Left) Measured K-band Strehl ratio vs H-band magnitude of the natural guide star obtained with the NIR PyWFS loop closed during commissioning from September 2018 to June 2019. (Middle) A typical K-band PSF with 60\% Strehl with the PyWFS loop closed. (Right) A very red target demonstrating the advatange of a NIR WFS.}
\end{figure}

The PyWFS first successfully closed the loop in November 2018. Commissioning continued to April 2018 when the PyWFS began preliminary science operations. Figure~\ref{fig:pywfsresults} summarizes the performance of the PyWFS over the first year of operation. The left panel shows the K-band Strehl ratio as a function of the H-band magnitude. The date when the data was collected is represented by the color of the dots with bluer dots corresponding to older data and redder dots more recent data sets. The shaded blue region highlights the range of expected Strehl ratios given a range of seeing conditions and PyWFS settings. It is clear that the majority of the Strehl ratio  measurements lay inside the blue shaded region, demonstrating the PyWFS is operating successfully. The redder and hence newer dots are almost always above the bluer older dots at any given magnitude emphasizing that progress was made over the course of commissioning in regards to optimizing and tuning the performance of the PyWFS. The Strehl was $\sim60\%$ in K-band down to 8$^{th}$ magnitude where it begins to roll off. AO correction has been achieved down to a H-band magnitude of 12, and could go as low as 13$^{th}$ which is promising for studying red objects.    
Upgrades to the system planned for the coming year will aim at increasing the throughput to the PyWFS, extending the operation to fainter stars.

The middle and right hand panels of~\ref{fig:pywfsresults} show some representative images taken with the PyWFS. The middle panel shows a typical $60\%$ K-band Strehl beam on a H=6.2 magnitude star (just above the roll off). The PSF clearly reflects the diffraction pattern of the Keck pupil indicating the stable and high quality correction provided by the AO system. The right panel shows a diffraction-limited PSF for the two components of a binary system that has a H=9.8 and a V$\sim15$ magnitude. This image clearly demonstrate the ability of the NIR PYWFS to provide diffraction-limited images on very red targets as the visible Shack-Hartmann at Keck would not be able to achieve such an image without the laser guide star mode.

In addition to quantifying the performance during commissioning, the PyWFS has been used for science observations with NIRC2 and the L/M-band vortex coronagraph.  These have demonstrated the superior performance of the the PyWFS in terms of contrast, with gains between 2 - 3 over the correction band compared to the Shack-Hartmann sensor.  The starlight suppression with the vortex is also improved when using the PyWFS, by typically $\sim50\%$. Further analysis of the PyWFS performance is reported in Bond et. al (2019)~\cite{Bond19}.

Now that the PyWFS is achieving the expected performance it will be facilitized and open up to the broader community.  This will include several hardware upgrades, including replacement of optics for higher throughput and the motorization of field steering mirrors to enable off-axis science. In addition, there are plans to implement predicative control~\cite{eden2019} and fine tune the overall performance of the AO system which should further enhance the performance.

\subsection{The fiber injection unit}
The FIU is designed to inject light into a SMF and maintain optimum coupling throughout a typical observing sequence (1-4 hrs). To achieve this a series of careful calibrations are required. 

The fiber bundle consists of several SMFs. Five SMFs are ZBLAN fibers that are single mode across K and L bands. Of these fibers, one fiber is designed to be positioned on the location of the planet, the three nearest to that are used to sample the stellar spectrum and the furthest one ($\sim3$~arcsec away) is designed to collect the background. The planet fiber will have residual starlight contaminating the signal so the aim is to use the fibers which only collect the stellar spectrum to be able to carefully calibrate the star out. In addition, the spectra collected by all fibers will be effected by tellurics from the atmosphere, equally, given the small field of view of the system, and the hope is that by measuring it contemporaneously we can calibrate this out as well. 

Besides the five science fibers, there are six other silica based fibers in the bundle. These fibers are single mode at 1550 nm and are used to reverse inject light from a 1550 nm laser source to determine the position of the bundle on the tracking camera carefully (see the right panel of Fig.~\ref{fig:lyot}). It has been determined that the bundle position is extremely stable with respect to the tracking camera so the location of the bundle on the camera only needed to be measured once. 

With the position of the fibers/bundle on the camera known, a fiber source internal to Keck AO is used to prepare the system before night time observing. The position of the fiber light source is noted on the tracking camera and the TTM is moved to steer the fiber source to the known location of the primary science fiber. To optimize the coupling, the TTM is scanned (in a raster pattern currently but other patterns will be used in future) while the slit viewing camera inside NIRSPEC (SCAM) is used to measure the flux through the fiber. In this way, a coupling map is created, which is used to locate the point of optimum coupling. The TTM is then steered to optimize the coupling. 

The next step is to minimize the non-common path errors (NCPA) between the PyWFS and the FIU. This is done through a two step process. The first step involves taking out of focus (applied by the Xinetics DM) images on the tracking camera and reconstructing the wavefront error using phase diversity techniques. This improves the Strehl at the location of the tracking camera, but there are additional NCPAs between the tracking camera and the fiber focal plane. To calibrate these, the Xinetics DM is used to apply the first 10 or so Zernike polynomials, one at a time and scan through the amplitude while monitoring the transmitted flux on SCAM. This algorithm has been written but is currently untested, but should help peak up the coupling by eliminating NCPA in the focal plane that matters. 

Once on-sky, the primary star in the system under investigation is tracked and its location shifted with the TTM to the position of the science fiber as determined by daytime calibrations. Given that the Keck AO system does not have an atmospheric disperion compensator, an extra calibration step must be implemented on-sky. Recall that the tracking camera operates in J and H band currently while the injection is at K and L and as such if the H-band PSF is well aligned with the pixel on the tracking camera that corresponds to the location of the fiber, the K or L-band PSF may be offset if observing at lower elevations. As such, the expected amount of differential chromatic dispersion between H and K or L is computed and the PSF is shifted in the tracking camera plane along the elevation axis by this amount. Because this relies on a very accurate model of atmospheric refraction which needs to take into account environmental properties such as temperature, pressure and humidity, there will always be some error to this. To correct for this, a scan of the TTM is once again performed on-sky using the SCAM detector to adjust the fine pointing of the TTM and maximise the flux on sky. At this point the primary is well aligned and now its time to inject the secondary (a binary, brown dwarf or a planet). 

Given that exoplanets are too faint to directly see on the tracking camera (and indeed so are some brown dwarfs), the parallactic angle and separation are acquired by previous knowledge of the state of the system and these are input into the control loop to shift the known companion to the location of the primary science fiber. This is a blind offset and requires a good distortion correction otherwise errors of 5-20 mas are possible causing a reduction in coupling, which is extremely critical for faint exoplanets. At this point the loop continues to track the location of the primary and maintain it while spectra of the companion can be collected. 

Most of this procedure has been developed and tested on-sky throughout 2019. The team is working on carefully calibrating the distortions in the tracking camera focal plane very carefully to enhance coupling but this should be operational by late 2019. 

\begin{figure} [t!]
   \begin{center}
   \begin{tabular}{c} 
   \includegraphics[width=0.98\textwidth]{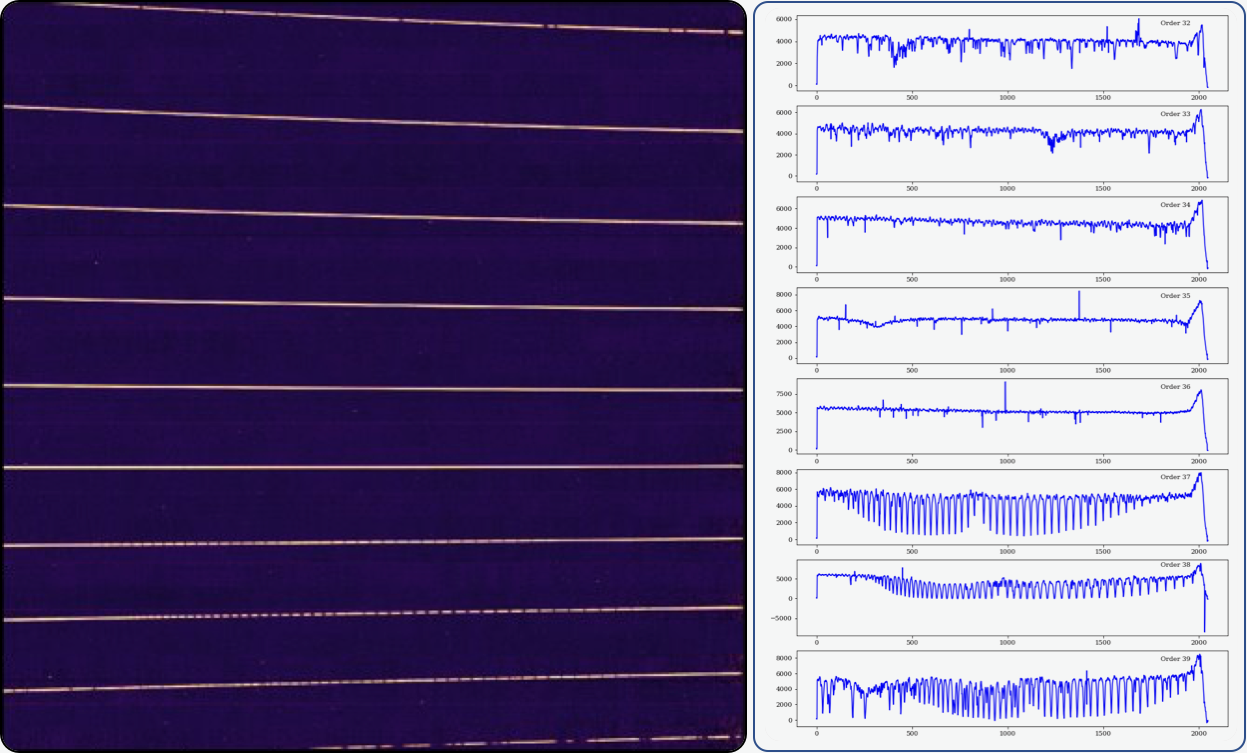}
   \end{tabular}
   \end{center}
   \caption[example] 
   { \label{fig:fiuresults} 
(Left) An image of the spectrum collected on the NIRSEC detector after dark subtraction and flat fielding for the star V* VR Ari. (Right) The crudely extracted spectrum of the star for each order clearly showing the presence of CO features in the bottom 3 orders.}
\end{figure}

Throughout the various engineering nights provided for FIU commissioning, the acquistion sequence has been tested and preliminary spectra collected. Figure~\ref{fig:fiuresults} shows one such data set. The spectra was collected on the star named V* VX Ari which has a M3V spectral typing and is $6^{th}$ magnitude in H and K bands. A spectrum was collected in K band and is shown in the left panel (after dark subtraction and flat fielding). The right hand panel shows the crudely extracted spectrum (note the bottom axis is simply pixel number and the vertical counts in ADU). In the bottom three orders its clear to see that there are a forest of narrow absorption features which correspond to the CO molecular spectrum. A proper data extraction pipeline which takes into account the recent NIRSPEC upgrades is currently under development, but these results nonetheless demonstrate the ability to inject a target into the SMF, feed it to the seeing limited instrument NIRSPEC and extract a high resolution spectrum.

The throughput of the system was measured to be $\sim1\%$. This is a lower than the expected $2.4\%$ we would expect from the phase I system. We believe that three effects led to this: 1) a misalingment between the star/fiber when the experiment was carried out, 2) residual NCPAs at the location of the FIU because we haven't commissioned the Zernike mode correction tool yet and 3) an inaccurate understanding of the throughput of NIRSPEC. We hope to remedy some of these over the next month and get the throughput towards $2\%$ which is sufficient to start HDC science.

\section{Summary}~\label{sec:summary}
The KPIC instrument will enhance exoplanet characterization capabilities at Keck. Phase I has been deployed and is nearing the end of commissioning while phase II is nearing the end of the design phase and rapidly moving to integration and testing. Phase I has demonstrated that stable high performance AO correction can routinely be achieved with the PyWFS on-sky enabling imaging of much redder objects than possible with the visible Shack-Hartmann sensor. The FIU is also operational and is capable of efficiently coupling and guiding a star on a SMF during an observing sequence. 

KPIC is now open for shared risk science and offers a unique platform for exploration and characterization. With phase II, the performance of the PyWFS and FIU will be enhanced enabling more efficient and more ambitious observations to be undertaken. 

The techniques and technologies developed through this unique project will lay the ground work for Earth-like planet characterization on future extremely large telescopes.

\acknowledgments 
 
The authors would like to acknowledge the financial support of the Heising-Simons foundation. We thank Dr. Rebecca Jensen-Clem for loaning AOSE for use within the KPIC phase II testbed. We would like to thank the SCExAO project for lending KPIC roof prisms for the NIR PyWFS. Part of this work was carried out at the Jet Propulsion Laboratory, California Institute of Technology, under contract with the National Aeronautics and Space Administration (NASA). The authors wish to recognize and acknowledge the very significant cultural role and reverence that the summit of Maunakea has always had within the indigenous Hawaiian community. We are most fortunate to have the opportunity to conduct observations from this mountain.

\bibliography{report} 
\bibliographystyle{spiebib} 

\end{document}